\documentclass[reprint, amsmath,amssymb, aps, showpacs, pra]{revtex4-1}
\usepackage{graphicx}
\usepackage{mathtools}
\usepackage{tensor}
\usepackage{diagrams}
\usepackage{verbatim}
\usepackage[breaklinks=true,colorlinks=true,linkcolor=blue,urlcolor=blue,citecolor=blue]{hyperref}
\usepackage{xcolor}
\usepackage{mdframed}
\usepackage{dsfont}

\def\expect#1{\left\langle #1 \right\rangle}
\def\bra#1{{\langle #1 |}}
\def\ket#1{{| #1 \rangle}}

\newcommand{\C}{\textsc{C}}

\begin{document}
\preprint{}

\title{Quantum information with general quantum variables: a formalism encompassing qubits, qudits, and quantum continuous variables}
\author{Timothy J. Proctor}
\thanks{Current address: Sandia National Laboratories, Livermore, CA 94550, USA}
\thanks{tjproct@sandia.gov}
\affiliation{School of Physics and Astronomy, E C Stoner Building, University of Leeds, Leeds, LS2 9JT, UK}

\begin{abstract}
This note presents a simple and unified formulation of the most fundamental structures used in quantum information with qubits, arbitrary dimension qudits, and quantum continuous variables. This  \emph{general quantum variables} construction provides a succinct language for formulating many results in quantum computation and information so that they are applicable in all dimensions. The structures included within this formalism include: a generalization to arbitrary dimension of the three Pauli operators, and the associated mutually unbiased bases; the Pauli and Clifford groups; many important quantum gates; standard sets of generators for the Clifford group; and simple universal gate sets. This formalism provides a convenient, intuitive and extensible language for easily generalizing results that were originally derived for a single type of system (often qubits or quantum continuous variables), and that rely on only those structures listed above, to apply in all dimensions. 
\end{abstract}
\maketitle

\noindent
{\bf Introduction.} The majority of research on quantum information and computation has focused on 2-dimensional quantum systems, i.e., qubits. There are good reasons for this: classical computers are built out of bits, the theory is arguably simplest for qubits, and experimental qubit-based devices are significantly more developed, by most metrics, than those for higher dimensional systems. However, there are also reasons to consider building quantum information devices out of higher dimensional \emph{qudits} ($d$-level systems, $d>2$) or  \emph{quantum continuous variables} (QCVs), which are systems with a single continuous degree of freedom. For example, there are interesting results showing that fault-tolerance thresholds for $d$-dimensional qudits are improved by increasing $d$ \cite{watson2015qudit,campbell2014enhanced,anwar2014fast,campbell2012magic,andrist2015error,duclos2013kitaev},  and optical QCVs are one of the easiest quantum systems to accurately control \cite{ukai2011demonstration,su2013gate,jensen2011quantum,yokoyama2013ultra}. 

Motivated by the potential of non-qubit systems -- and because so much of the quantum information literature focuses exclusively on qubits or QCVs -- there is a substantial body of literature on extending qubit-based results to apply to qudits \cite{zhou2003quantum,hostens2005stabilizer,Proctor2015ancilla,proctor15measurement,Proctorthesis2016,sheridan2010security,zilic2007scaling,howard2012qudit,
parasa2011quantum,cao2011quantum,proctor2014quantum,joo2007one,gottesman1999fault,
farinholt2014ideal,garcia2013swap} or QCVs \cite{menicucci2006universal,braunstein2005quantum,gu2009quantum,proctor15measurement,bartlett2002efficient,
Proctor2015ancilla,Proctorthesis2016}, or converting from QCVs to qubits \cite{Proctorthesis2016,proctor2014quantum}. One purpose of this paper is to make the process of translating results from one type of system into another easier, in those cases in which the results rely on only those structures which are common to qubits, arbitrary dimension qudits, and QCVs. Specifically, this paper introduces the framework of the \emph{general quantum variable} (GQV), an entity of unspecified dimension: it is parameterized by a dimension $d$ that may correspond to either a qubit, a qudit of any dimension, or a QCV.

To my knowledge, the formalism presented here is novel. However, every result presented herein is not novel in at least the qubit case, and most are not novel in any dimension. This work should be thought of as a framework for dimension-independent quantum information theory. Moreover, this paper is \emph{not} intended to be  an all-encompassing review of what is known about QCVs, qudits and qubits. In particular, there are many  properties of quantum systems which \emph{do} depend strongly on some property of the dimension, and so will not be included herein. For example, many results are built on the structures of a field, and so only apply for prime dimension qubits, QCVs, and sometimes power-of-a-prime dimension qudits \cite{watson2015qudit,campbell2014enhanced,anwar2014fast,campbell2012magic,andrist2015error,duclos2013kitaev,wootters1987wigner,vourdas2003factorization,gibbons2004discrete,klimov2005multicomplementary,klimov2009discrete,silberhorn2007detecting}.

\vspace{0.2cm}
\noindent
{\bf General quantum variables.} A $d$-dimensional \emph{qudit} is a quantum system with a Hilbert space of finite dimension $d\in \mathbb{N}$ for $d\geq 2$ \cite{vourdas2004quantum,weyl1950theory}. The particular case of $d=2$ is a qubit. The integers modulo $d$ plays a crucial role for qudits, which is the set $\{0,1, \dots, d-1\}$ equipped with modular arithmetic \cite{vourdas2004quantum}. This structure is a \emph{ring} -- because division is not generally well-defined -- and it is denoted by $\mathbb{Z}(d)$ herein.  

A quantum continuous variable (QCV) is a quantum system with a single continuous degree of freedom taking values in $\mathbb{R}$ \cite{Lloyd1999quantum,braunstein2005quantum}. Equivalently, a QCV is a quantum system described by one-dimensional wave mechanics with ``position'' and ``momentum'' operators, denoted $\hat{x}$ and $\hat{p}$ respectively, obeying the canonical commutation relation
\begin{equation}
[\hat{x},\hat{p}]=i.
\end{equation}

A \emph{general quantum variable} (GQV) is defined to be a quantum system parameterized by a dimension-signifier $d$ where $d=d' \in \mathbb{N}_{\geq 2}$ to obtain the case of a $d'$-dimensional qudit, and $d= 2\pi$ to obtain the case of a QCV (this $2\pi$ convention is chosen as it results in succinct notation, as $2\pi$ has a fundamental significance for QCVs). The underlying structures on which a $d$-dimensional qudit and a QCV are defined are different, so it is notationally useful to define the ring $\mathbb{S}_d$ by
\begin{equation}
\mathbb{S}_d :=  \begin{cases}
    \mathbb{Z}(d)       & \quad \text{for a } d\text{-dimensional qudit},  \\
\mathbb{R}  & \quad\text{for a QCV}. \\
  \end{cases}
\end{equation}

For qudits the group of all the $n$-qudit unitaries, which is denoted herein by $\mathcal{U}(d^n)$, is important in quantum computation theory. For QCVs, the relevant group of transformation is more subtle. It is conventional to only consider the subset of $n$-QCV unitaries containing all operators of the form 
\begin{equation}
 \mathcal{U}_{n} := \{  U \mid U=\exp( i \text{poly}(\hat{x}_k,\hat{p}_k))\}.
\end{equation}
 where $\text{poly}(\hat{x}_k,\hat{p}_k)$ is any finite-degree polynomial (over $\mathbb{R}$) of the position and momentum operators of all $n$ QCVs \cite{proctor15measurement,Lloyd1999quantum}. For notational simplicity, denote this set by 
\begin{equation}
\mathcal{U}((2\pi)^n) \equiv \mathcal{U}_n, 
\end{equation}
so that for a GQV the relevant set of unitaries in quantum computation is denoted $\mathcal{U}(d^n)$.

So far, nothing significant has been achieved -- notation has being used to hide the differences between qudits and QCVs. However, now that this notation has been defined, the critical structures for qubit, qudit and QCV quantum information theory can be presented so as to make the equivalences between standard quantum information structures in different dimensions particularly clear.

\vspace{0.2cm}
\noindent
{\bf The computational basis.} A common starting point for quantum computation theory is to introduce a computational basis, from which all other structures are defined (the alternative is to start with observables and use these to define this basis). For all types of GQVs, some computational basis of the form
\begin{equation}
\mathcal{B}:=\{\ket{q} \mid q \in \mathbb{S}_d \},
\end{equation}
 may be chosen for the relevant Hilbert space, with the basis states obeying 
\begin{equation}
\expect{q|q'} = \delta(q-q').
\end{equation}
Here $ \delta(\cdot)$ denotes the Dirac delta function in the case of QCVs \cite{sakurai1985modern} and the Kronecker delta function in the case of qudits. For example, in the case of QCVs this basis may be defined as the generalized eigenstates of $\hat{x}$. Note that these states are unphysical for QCVs, but they may be approximated to arbitrary accuracy with a physical state \cite{braunstein2005quantum}\footnote{To be more exact, these states are not part of the Hilbert space of square integrable functions and so it is necessary to employ the larger structure of a ``rigged Hilbert space'' to consider these states in a mathematically precise manner \cite{de2005role}. These technicalities are normally ignored in the theory of QCV quantum computation, but note that rigged Hilbert spaces provide a solid basis for QCV manipulations using these states. These states can be approximated to any desired precision with a physically well-defined state. For example, they may be approximated by a state with a Gaussian wave function centered on $q$ with a narrow peak, which is a squeezed state \cite{braunstein2005quantum}.}.

Later we will require the definition for mutually unbiased bases in all dimensions. A set of orthonormal bases $\{\mathcal{B}_1,\mathcal{B}_2,...\}$ for a GQV are called \emph{mutual unbiased} if for any pair of bases $\mathcal{B}_j$ and $\mathcal{B}_k$ in this set, with $j\neq k$, and for any basis states $\ket{a} \in \mathcal{B}_j$ and $\ket{b} \in \mathcal{B}_k$ in these bases,
\begin{equation} 
| \langle a | b \rangle |^2  = k_d ,
\end{equation}
where $k_d$ is any non-zero and positive constant \cite{weigert2008mutually,durt2010mutually}. To satisfy this relation for qudits it is necessary for $k_d=1/d$ \cite{durt2010mutually}.

\vspace{0.2cm}
\noindent
{\bf The Fourier gate and basis.} Starting from the computational basis, we can define the \emph{Fourier gate} ($F$). This is the unitary representation of the appropriate dimension Fourier transform. Specifically,
\begin{equation}
 F\ket{q}:= \frac{1}{\sqrt{d}} \sum_{q' \in \mathbb{S}_d} \omega^{qq'} \ket{q'},
 \end{equation}
with 
\begin{equation}
\omega : = \exp(2 \pi i /d).
\end{equation}
 The $\sum_{q' \in \mathbb{S}_d}$ notation denotes that the summation is over all values in $\mathbb{S}_d$, e.g., it is an integral over $\mathbb{R}$ for QCVs. It is simple to show that 
\begin{equation}
F^4 = \mathbb{I},
\end{equation}
and that $F$ is unitary (see Appendix~\ref{App:Fourier-gate}). For qubits, the Fourier gate is the well-known Hadamard gate, normally denoted $H$. For QCVs, the Fourier gate is generated by the quantum harmonic oscillator Hamiltonian \footnote{Specifically, for a QCV, $F$ is generated by the Hamiltonian$\hat{H}_{\textsc{qho}} = \frac{1}{2} \left( \hat{x}^2 + \hat{p}^2 \right)$ applied for a time $t=3\pi/2$.}.

Using the Fourier gate, a \emph{Fourier basis} $\mathcal{B}_{+}$ (the ``conjugate basis'' is an alternative well-motivated name) can be defined by the set of orthonormal states 
\begin{equation}
\ket{+_q} := F \ket{q},
\end{equation}
 for $ q \in \mathbb{S}_d$. This notation is adapted from that in common usage for qubits, where conventionally $\ket{+} \equiv \ket{+_0}$ and $\ket{-} \equiv \ket{+_1}$. It is simple to confirm that
   \begin{equation}
    \expect{q | +_{q'}} = \frac{ \omega^{qq'}}{\sqrt{d}}  \hspace{0.5cm} \forall \, q,q' \in \mathbb{S}_d. \label{overlapcomcon}
    \end{equation}
This follows because each conjugate basis state is an equal superposition (with phase factors) of all possible computational basis states. This relation implies that the computational and Fourier bases form a set of two mutually unbiased bases.

\vspace{0.2cm}
\noindent
{\bf The phase gate and basis.} By introducing an additional gate (that is also important in its own right) we may define a third basis that is, arguably, equally as fundamental as the Fourier and computational bases. Define the parameterized \emph{phase gate} by 
\begin{equation} 
P(p) \ket{q}:= \omega^{\frac{pq}{2}(q+\varrho_d )} \ket{q}, \label{phasegate}
 \end{equation}
with $p\in\mathbb{S}_D$, where 
\begin{equation}
\mathbb{S}_D :=  \begin{cases}
    \mathbb{Z}(2d)      & \quad \text{for a } d\text{-dimensional qudit},  \\
\mathbb{R}  & \quad\text{for a QCV}, \\
  \end{cases}
\end{equation}
and
\begin{equation}
\varrho_d:=  \begin{cases}
    1      & \quad \text{for odd-dimension qudits },  \\
0 & \quad\text{otherwise}. \\
  \end{cases}
\end{equation}
The $d$-dependent $\varrho_d$ parameter is not essential in the phase gate definition. However, it is useful for reducing the odd/even dimension dependence of the properties of this gate. For qubits the phase gate reduces to the well-known gate $P= \ket{0}\bra{0}+i \ket{1} \bra{1}$, often also denoted by $S$. 

Using the phase gate, define the \emph{phase basis} by
\begin{equation} 
\mathcal{B}_{\times} := \{ \ket{\times_q} := PF \ket{q} \mid q \in \mathbb{S}_d \}, 
\label{Eq_phasebasis}
\end{equation}
where, here and throughout, $G$ denotes $G(1)$ for a parameterized gate $G(\cdot)$. That is $P \equiv P(1)$. The computational, Fourier, and phase bases are particularly useful because they form a set of three mutually unbiased bases. In particular, they satisfy
\begin{align} 
    \expect{q | +_{q'}} &= \frac{ \omega^{qq'}}{\sqrt{d}},\\
\expect {q| \times_{q'}}  &= \frac{ \omega^{ qq'}}{\sqrt{d}}    \omega^{-\frac{q}{2} (q+\varrho_d)} \label{overlapcompha} ,\\ 
  \expect{+_q  | \times_{q'}} & =\frac{  \omega^{qq'}}{\sqrt{d}}  \omega^{-\frac{q}{2}(q-\varrho_d)}  \omega^{-\frac{q'}{2}(q'+\varrho_d)}   \omega^{\frac{d-\varrho_d}{8} }  \label{overlapconpha},
\end{align}
for all $q,q' \in \mathbb{S}_d$. The first of these equations has already been stated in Eq.~\eqref{overlapcomcon}, and deriving it is simple. The latter two of these equations are derived in Appendix~\ref{Amub}. The computational, Fourier and phase basis states have another important property: they are the eigenstates of perhaps the most natural generalization of the Pauli operators to arbitrary dimension.

\vspace{0.2cm}
\noindent
{\bf The Pauli operators.} The Pauli operators are the most fundamental structure for a qubit, and are defined by 
\begin{align}
\sigma_x &=\ket{1}\bra{0}+\ket{0}\bra{1},\\
i\sigma_y &=\ket{0}\bra{1}- \ket{1}\bra{0},\\
\sigma_z &=\ket{0}\bra{0}-\ket{1}\bra{1}.
\end{align}
The natural (and well-known) \emph{unitary} generalization of the $\sigma_x$ and $\sigma_z$ Pauli operators to arbitrary dimension are the $q' \in \mathbb{S}_d$ parameterized unitaries defined by
\begin{equation} 
Z(q') \ket{q}  := \omega^{qq'} \ket{q} , \hspace{1cm} X(q') \ket{q}  := \ket{q+q'} ,  \label{Paulicomp} 
\end{equation}
for all $q,q' \in \mathbb{S}_d$. In these definitions the arithmetic should be taken to be that of $\mathbb{S}_d$, as should be assumed for all arithmetic throughout unless otherwise stated. For QCVs, these are often termed position and momentum translations, and they are equivalent to the ``displacement operator'' along two orthogonal axes in phase space (see Appendix~\ref{sec:QVs}). For finite dimension, and outside of the quantum computation literature, these operators are sometimes known as the ``phase'' and ``clock'' matrices, respectively.

In the literature, it does not seem to be conventional to also generalize the qubit $\sigma_y$ gate to other dimensions. However, there is a gate that shares the core properties of $\sigma_y$, defined by:
\begin{equation}
Y(q):=\omega^{q(q+\varrho_d)/2}Z(q)  X(q).
\end{equation}
To be more consistently with the definitions for the $X$ and $Z$ Pauli operators used herein, this gate could equivalently be defined by its action on the computational basis (which is given below). For the case of a qubit, $X\equiv X(1)$, $Y\equiv Y(1)$, and $Z \equiv Z(1)$ are equal to $\sigma_x$, $\sigma_y$, and $\sigma_z$, respectively.

The Fourier, phase, and computational bases are the eigenstates of the $X(q)$, $Y(q)$ and $Z(q)$ gates, respectively, with eigenvalues that are powers of $\omega$. Each of the three GQV Pauli operators permutes the eigenstates of the other two GQV Pauli operators. The complete set of relations is
\begin{align}
X(q') \ket{q}  &= \ket{q+q'},\label{eq:Xc} \\
Y(q') \ket{q}  &= \omega^{q'(3q'+2q+\varrho_d)/2}\ket{q+q'},\label{eq:Yc}\\
Z(q') \ket{q}  &= \omega^{qq'} \ket{q},\label{eq:Zc}\\
 X(q') \ket{+_{q}} &= \omega^{-qq'} \ket{+_q},\label{eq:Xf} \\
 Y(q')\ket{+_{q}} &=\omega^{q'(q'-2q+\varrho_d)/2} \ket{+_{q+q'}},\label{eq:Yf}  \\
 Z(q') \ket{+_{q}} &= \ket{+_{q+q'}}, \label{eq:Zf}\\
 X(q')\ket{ \times_q} &=\omega^{q'(q'-2q-\varrho_d)/2}  \ket{\times_{q-q'}},\label{eq:Xp}\\
Z(q') \ket{ \times_q} &= \ket{\times_{q+q'}}, \label{eq:Yp}\\
Y(q') \ket{ \times_q} &=\omega^{-qq'}  \ket{\times_{q}}, \label{eq:Zp},
\end{align}
for all $q,q' \in \mathbb{S}_d$. Eqs.~\eqref{eq:Xc} and \eqref{eq:Zc} hold by definition, and Eq.~\eqref{eq:Yc} follows from these two equations and the definition of $Y(\cdot)$. The Fourier basis relations of Eqs.~(\ref{eq:Xf} -- \ref{eq:Zf}) follow easily from the definition of the Fourier transform. The actions of the three Pauli operators on the phase basis, given in Eqs.~(\ref{eq:Xp} -- \ref{eq:Zp}), are proven in Appendix~\ref{Amub}.

\vspace{0.2cm}
\noindent
{\bf Hermitian position and momentum operators.} In the context of QCVs, it is common to define most quantities in terms of the position and momentum operators. For GQVs we may define the Hermitian ``position'' and ``momentum'' operators by
 \begin{equation} 
 \hat{x} := \sum_{q\in\mathbb{S}_d} q\ket{q}\bra{q},\hspace{1cm} \hat{p} := \sum_{q\in\mathbb{S}_d} q\ket{+_q}\bra{+_q}.
 \end{equation}
The Pauli operators can also be expressed as exponentials of $\hat{x}$ and $\hat{p}$, specifically
\begin{equation}
  X(q)=\omega^{-q\hat{p}}, \hspace{0.5cm} Z(q)=\omega^{q \hat{x}} , \hspace{0.5cm} q \in \mathbb{Z}_d.
  \label{eq:QCVpauli-xp}
 \end{equation}
These equations are often chosen to define the Pauli gates in the QCV setting.

\vspace{0.2cm}
\noindent
{\bf The Pauli group.}
An important property of the Pauli operators is that they commute up to a phase. This is summarized by the relation
\begin{equation} 
Z(q) X(q') = \omega^{qq'} X(q') Z(q), \label{Weyl} 
\end{equation}
 for all $q,q' \in \mathbb{S}_d$, which may easily be confirmed by the action of each side of the equality on the computational basis. This is often called the \emph{Weyl commutation relation}. From the Weyl commutation relation it follows that tensor products of the Pauli operators generate a subgroup of the $U(d^n)$, called the Pauli group.

 The \emph{Pauli group} ($\mathcal{P}$) is defined to consist of all operators of the form
\begin{equation}
 p_{\xi, \vec{q}} : = \omega^{\xi/2} X(q_1) Z(q_{n+1})  \otimes  .... \otimes X(q_{n}) Z(q_{2n}) ,
 \end{equation}
where 
\begin{equation} \vec{q}=(q_1,\dots,q_{2n}) \in \mathbb{S}_d^{2n},
\end{equation}
 and $\xi \in\mathbb{S}_D$. This reduces to the well-known qubit Pauli group for $d=2$; for a single qubit this is the group with elements
\[ \mathcal{P}_{\text{qubit}} = \{\pm\mathds{1},\pm i \mathds{1},\pm \sigma_x,  \pm i \sigma_x, \pm\sigma_y,\pm i\sigma_y,\pm\sigma_z,\pm i\sigma_z\},
\]
where $\mathds{1}$ is the identity \cite{gottesman1999heisenberg}. For QCVs, the Pauli group is often called the Heisenberg-Weyl group \cite{bartlett2002efficient}. Pauli operators compose as
\begin{equation}
p_{\xi,\vec{q}} \, p_{\zeta,\vec{p}}=p_{\xi+\zeta+2\delta,\vec{q}+\vec{p}},
\end{equation}
where\
\begin{equation}
\delta = \sum_{i=1}^{n} q_{i}p_{n+i}. 
\end{equation}
If the GQV is a $d$ dimensional qudit, it is perhaps ambiguous as to whether to calculate $\delta$ using modulo $d$ or $2d$ arithmetic. However, $\omega^{\delta}$ is invariant under changing this convention, so the choice is irrelevant.

\vspace{0.2cm}
\noindent
{\bf The Clifford group.}
The Clifford group is of fundamental importance in quantum computation, for example, it underpins much of the theory of quantum error correction \cite{gottesman2010introduction}. The $n$-GQV Clifford group is the normalizer of the Pauli group in the group of the unitaries. That is, it is defined by \cite{gottesman1999heisenberg,bartlett2002efficient,gottesman1999fault}
\begin{equation}
 \mathcal{C}: = \{ U \in U(d^n) \mid U  p U^{\dagger} \in \mathcal{P} \hspace{0.2cm} \forall   p  \in \mathcal{P} \}. \end{equation}
The Clifford gates are the unitaries that transform Pauli gates to Pauli gates under conjugation. Note that Clifford gates are normally referred to as ``Gaussian operations'' in the setting of QCVs \cite{braunstein2005quantum}. 

The Fourier, phase and Pauli operators are Clifford gates. The Fourier gate transforms cyclically between $X(q)$ and $Z(q)$. Specifically, under conjugation by the Fourier gate, the Pauli operators are transformed with the relation
\begin{equation}
\begin{diagram}  X(q) & \rTo & Z(q) \\ \uTo & & \dTo \\ Z(-q) & \lTo & X(-q) \end{diagram} \label{Ftrans}  \end{equation} 
where this represents the relation that $F X(q) F^{\dagger} = Z(q)$, and so on.  

Another useful and commonly encountered single GQV Clifford operator is the \emph{squeezing gate} defined by
\begin{equation} 
S(s) |q \rangle :=  | sq \rangle , 
\label{squeezing-gate-def}
\end{equation}
for any $s\in\mathbb{S}_d$ such that there exists $s^{-1} \in\mathbb{S}_d$, where $s^{-1}$ is defined to be the element of $\mathbb{S}_d$ that satisfies $s^{-1}s = 1$, if such an element exists \footnote{This condition holds for all non-zero $s\in\mathbb{S}_d$ with QCVs and prime dimension qudits, and any non-zero $s$ which is co-prime with the dimension of the qudits in all other cases (such an $s$ is called a unit in $\mathbb{S}_d$).}. Note that this condition  on the argument of $S(\cdot)$ is required for the squeezing gate to be unitary, as when this is violated the transformation cannot be inverted.

Perhaps the three most important two-GQV Clifford gates in quantum information theory are:
\begin{align} 
\textsc{cz}\ket{q}\ket{q'}&:= \omega^{qq'}\ket{q}\ket{q'},\\
\textsc{sum}  \ket{q}|q' \rangle &:= \ket{q}| q+q' \rangle ,\\
\textsc{swap} \ket{q}|q'\rangle &:= |q'\rangle \ket{q},
\end{align}
The first two of these gates are entangling, and one can be converted to the other by conjugating the second QGV with Fourier gates. The \textsc{sum} gate is conventionally called \textsc{cnot} in the case of qubits.  As the name suggests, the action of \textsc{swap} is to swap the states of the two input GQVs.


\vspace{0.2cm}
\noindent
{\bf Generating the Clifford group.} The $\textsc{cz}$, $F$, $P(p)$ and Pauli gates form a set of generators for the Clifford group \cite{hostens2005stabilizer,farinholt2014ideal,bartlett2002efficient}. That is:
\begin{equation}
\mathcal{C} = \langle \textsc{cz},F, P(q), Z(q) \rangle \,\,\,\,\text{with}\,\,\,\, q\in \mathbb{S}_d, \label{Cliffgen} 
\end{equation} 
where this uses the standard group theory notation that $\mathcal{G} = \langle g_1,\dots,g_k \rangle$ represents the statement that $\mathcal{G}$ is the group generated by the elements $g_1,\dots,g_k$. Explicitly, any $n$-GQV Clifford gate can be exactly decomposed into multiplicative and tensor products of these four gates (two of which are parameterized). 

There exists a decomposition of an $n$-GQV Clifford that contains only $O(n^2)$ of these basic generating gates \cite{hostens2005stabilizer,farinholt2014ideal,bartlett2002efficient}. From some dimensions of QCV, not all of these generators are required in order to generate the Clifford group. For example, $Z(q)$ and $P(q)$ can be obtained from integer powers of $Z$ and $P$ for any dimension qudit, reducing the number of generators to four constant (i.e., not parameterized) gates. Moreover, for a qubit or any odd-dimension qudit, $Z$ can be obtained from $P$ and $F$. Specifically, for qubits and odd dimension qudits $F^2P^{d-1}F^2P=Z$ \footnote{It is claimed in Ref.~\cite{farinholt2014ideal} that for qudits of \emph{all} dimensions the $Z$ generator can be removed from the generating set given here. However, I am unaware of an explicit proof, so this claim has been omitted here.}.

When performing manipulations with Clifford gates, it is often useful to have explicit expressions for how particular important Clifford gates transform Pauli gates. Letting $U \xrightarrow{  \mathmakebox[0.4cm]u} U'$ denote that $uUu^{-1}=U'$, we have that
\begin{align}
  p_{\xi, (q_1,q_2)}   &\xrightarrow{ \mathmakebox[1.2cm]{Z(p)}} p_{\xi+2pq_1,(q_1,q_2)}, \label{Eq:conj-Z} \\   
  p_{\xi, (q_1,q_2)}   &\xrightarrow{ \mathmakebox[1.2cm]{F}} p_{\xi-2q_1q_2,(-q_2,q_1)}, \label{Eq:conj-F} \\
 p_{\xi ,(q_1,q_2)}  &\xrightarrow{  \mathmakebox[1.2cm]{P(p)} } p_{\xi+pq_1(q_1+\varrho_d) ,(q_1,q_2+pq_1)} , \label{Eq:conj-P} \\ 
 p_{\xi ,q,q'}  & \xrightarrow{ \mathmakebox[1.2cm]{S(p)}} p_{\xi,pq,p^{-1}q'}, \\
p_{\xi ,(q_1 ,q_2 ,q_3,q_4)} &\xrightarrow{ \mathmakebox[1.2cm]{\textsc{cz}} }    p_{\xi +2q_1q_2,(q_1,q_2,q_3+q_2,q_4+q_1)} \label{Eq:conj-CZ}. \end{align}
These relations are proven in Appendix~\ref{App:Clifford}. Note that the arithmetic in each subscript is calculated as appropriate for each type of GQV. In particular, for qudits it is modulo $2d$ for the phase factor and modulo $d$ for the latter two indices.

\vspace{0.2cm}
\noindent
{\bf Universal quantum computation.}
Within the GQVs formalism, the standard notion of an approximately universal quantum computer is equivalent in all dimensions. Specifically, an $n$-GQV (approximately) universal quantum computer is a device which can approximate to arbitrary accuracy any unitary operator in $U(d^{n})$ on $n$ GQVs \cite{Lloyd1999quantum,brylinski2002universal}. Any two-GQV entangling gate along with a set of single-GQV gates that can approximate, to arbitrary accuracy, any single-GQV gate is universal \cite{Lloyd1999quantum,brylinski2002universal}. 

Conceptually simple universal gate sets can be constructed from Clifford gates and single-QGV gates that only add phases to the computational basis states. The \emph{computational basis rotation gate} $R(\vartheta)$  takes a function parameter, $\vartheta: \mathbb{S}_d \to \mathbb{R}$, and is defined by
\begin{equation}
 R(\vartheta)  \ket{q} := e^{ i \vartheta(q)} \ket{q} .\label{Zdt}
\end{equation}
For all types of GQV, some set of rotation gates along with the Fourier gate are a universal set for single-GQV gates \cite{proctor15measurement,Lloyd1999quantum,zhou2003quantum} and hence such a set along with an entangling gate is sufficient for UQC.

To guarantee that this rotation gate is well-defined for a QCV, the $\vartheta$ function should be constrained to being some finite-degree polynomial in $q$ (over the real numbers). There is a physically irrelevant global phase freedom in this gate, which can be removed by setting $\vartheta(0)=0$. In the case of qudits, the gate set of all such rotation gates \footnote{For qudits, these rotation gates are parametrized by $d$ phase angles in $\mathbb{R}$, or $d-1$ phase angles if the global phase is fixed} along with the Fourier gate is an exactly universal single-qudit set. This is well-known for qubits \cite{nielsen2010quantum}\footnote{Any single-qubit unitary may be written as $U=e^{i\phi} R(\theta)HR(\phi)HR(\gamma)$ where $R(\chi)\ket{q}=e^{i\chi q}\ket{q}$ and $H$ is the Hadamard gate, i.e., the qubit Fourier gate. This may be shown directly via simple matrix multiplication.} and for qudits it is implied by the results of Zhou \emph{et al.}~\cite{zhou2003quantum}. This can then be adapted to a finite gate set and approximate universality by simply picking a set containing one `generic' rotation gate along with the Fourier gate $F$ \cite{proctor15measurement}. This is shown in Appendix~\ref{Aunigen} and is again well-known for the qubit sub-case.

\vspace{0.2cm}
\noindent
{\bf Universality with the Clifford group + 1.}
Gates from the Clifford group are alone not sufficient for universal quantum computation, as they form a strict subset of the unitary group. Furthermore, quantum computations consisting of preparing GQVs in the computational basis, applying a circuit of only Clifford gates, and then measuring the GQVS in the computational basis can be efficiently exactly simulated on a classical computer. This is generally known as the Gottesman-Knill theorem \cite{gottesman1999heisenberg, bartlett2002efficient,hostens2005stabilizer,gottesman1999fault,de2013linearized,van2013efficient}.

Yet, for \emph{prime dimension} qudits there is a particularly elegant result: the addition of \emph{any} non-Clifford gate to a set of generators of the Clifford group elevates that set to (approximate) universality \cite{campbell2012magic,nebe2006self,nebe2001invariants}. Similarly, for QCVs it is known that the addition of continuous powers of any non-Clifford single-QCV gate to the Clifford group is sufficient for (approximate) universality \cite{Lloyd1999quantum}. 

The non-Clifford gate often considered is a so-called \emph{cubic phase gate} of some sort, which can be defined in general by
\begin{equation} 
D_{3}(q') \ket{q} := \omega^{q^3q'/c} \ket{q}, 
\label{cubic-phase}
\end{equation}
for $q' \in \mathbb{S}_d$ and some suitable constant $c$. For all prime dimensions we can take $c=d^3$. This is the natural generalization of the well-known $T$ gate for qubits (which is a $\pi/4$ rotation around $\sigma_z$) \cite{Proctorthesis2016}. For prime $d>3$ qudits, $c=1$ also provides a non-Clifford gate \cite{campbell2014enhanced}, and for QCVs $c=3$ is conventional \cite{gu2009quantum} (note that in this case the value of $c$ is essentially irrelevant). For non-prime dimension qudits, and as already noted above, the addition of any $R(\vartheta)$ gate for a ``generic'' fixed $\vartheta$ to some Clifford group generators is sufficient for universality \cite{proctor15measurement}.

\vspace{0.2cm}
\noindent
{\bf Summary.} This paper has introduced \emph{general quantum variables} (GQV), as a simple and unified formulation of the most fundamental structures used in quantum information with qubits, qudits, and quantum continuous variables. This construction provides a succinct language for formulating many results in quantum computation and information theory so that they are applicable in all dimensions. The GQVs notation can streamline the process of translating known results between dimensions, and can facilitate the derivation of new results that are immediately applicable to qubits, qudits and QCVs.

\vspace{0.2cm}
\noindent
{\bf Acknowledgments.} I thank Viv Kendon, Paul Knott, Dan Browne, and Jiannis Pachos for helpful comments related to this work, and Birgitta Whaley for hosting me at the University of California Berkeley for a large portion of this work. The author was funded in part by a University of Leeds Research Scholarship.

\appendix
\bibliographystyle{apsrev}
\bibliography{Bib_Library} 
  
\section{The Fourier gate \label{App:Fourier-gate}}
In this appendix the properties of the Fourier gate ($F$) that are stated in the main text are derived. The following relation will be useful:
 \begin{equation} 
  \frac{1}{d} \sum_{r \in \mathbb{S}_d} \omega^{r (q-q')} = \delta(q-q'),    \label{Eq:sum-roots-delta} \end{equation}
  where $q-q'$ is taken modulo $d$ for qudits. For qudits this is straightforward to prove directly, using the formula for a geometric series. For QCVs it holds because the Fourier transform of a complex exponential function $e^{-iq'r}$ is a delta function, with the exact relation given by \cite{bateman1954tables}
\begin{equation}
  \frac{1}{\sqrt{2\pi} } \int^{\infty}_{-\infty} \text{d}r e^{-q' r} e^{+iqr} = \sqrt{2\pi} \delta(q-q').
\end{equation}

First it is shown that $F$ is a unitary operator. Using Eq.~\eqref{Eq:sum-roots-delta} and the orthogonality relation $\expect{q|q'}=\delta(q-q')$, it follows that
\begin{align}
 F F^{\dagger} &  = \frac{1}{d} \sum_{q,q' } \sum_{r,r'  }   \omega^{q q'-rr'} \ket{q}\expect{q' | r } \bra{r' }, \\
& = \frac{1}{d}\sum_{q} \sum_{r,r' }  \omega^{r(q-r')} \ket{q}\bra{r' },\\
& = \sum_{q} \sum_{r'  }   \delta(q-r')   \ket{q}\bra{r' },\\
& =  \sum_{q }   \ket{q}\bra{q}, \\
&= \mathbb{I},
\end{align}
with all summations over $\mathbb{S}_d$.
The same derivation holds to show that $F^{\dagger}F=\mathbb{I}$, confirming that $F$ is unitary. A minor alteration of this derivation shows that 
\begin{equation}
F^2 = \sum_{q \in \mathbb{S}_d} \ket{-q}\bra{q}.
\end{equation}
This is a useful relation in itself, and it implies that $F^4=\mathbb{I}$, as stated in the main text.

\section{The phase basis\label{Amub}}
In this appendix the action of the Pauli operators on the phase basis and the overlaps between the states in the computational, conjugate and phase bases are derived. This will include the proof that these bases are a set of three mutually unbiased bases for all types of QV. The phase basis is defined by
\begin{equation} 
\mathcal{B}_{\times} := \{ \ket{\times_q} := PF \ket{q} \mid q \in \mathbb{S}_d \}, 
\label{Eq_phasebasis-Ap}
\end{equation}
where $P$ has the action
\begin{equation} 
P\ket{q} = \omega^{q(q+\varrho_d)/2} \ket{q}.
\label{Eq_phasegate-Ap}
\end{equation}
Using the definition of the phase basis, the Pauli conjugation relation for the phase gate given in Eq.~\eqref{Eq:conj-P}, and the action of the Pauli operators on the Fourier basis given in Eqs.~(\ref{eq:Xf} -- \ref{eq:Zf}), it follows that
\begin{align*}
 \omega^{\xi/2}X(a)Z(b) \ket{ \times_q}   &= \omega^{\xi/2}X(a)Z(b)  P\ket{+_q},
 \\&=P \omega^{(\xi-a(a+\varrho_d))/2}X(a)Z(b-a)   \ket{+_q}  ,
 \\&=\omega^{(\xi-a(a+\varrho_d))/2+a(a-b-q)} \ket{\times_{q+b-a}},
  \\&=\omega^{(\xi+a(a-\varrho_d))/2-a(b+q)} \ket{\times_{q+b-a}}.
\end{align*}
This then implies the Eqs.~(\ref{eq:Xp} -- \ref{eq:Zp}).

Consider the overlap between an arbitrary phase basis state and an arbitrary computational basis state. Using the defining action of the phase gate on the computational basis and the overlap $\expect{q|+_{q'}} = \omega^{qq'}/\sqrt{d}$, it follows that
\begin{align} \expect {q| \times_{q'}} &= \bra{q} P\ket{+_{q'}} ,
\\&= \omega^{-q (q + \varrho_d )/2} \expect{q| +_{q'}} ,
\\&= \omega^{ q(q'- (q+\varrho_d )/2)} /\sqrt{d} ,
\end{align}
 for all $q,q' \in \mathbb{S}_d$, as stated in Eq.~\eqref{overlapcompha} of the main text. Now, consider the overlap of arbitrary conjugate and phase basis states. Again, using the action of the phase gate on the computational basis and the overlap of the conjugate and computational bases, it follows that
\begin{align} \expect {+_q| \times_{q'}} & =\vphantom{\frac{1}{d}} \bra{+_q} P\ket{+_{q'}},
 \\ & =\vphantom{\frac{1}{d}} \sum_{k\in\mathbb{S}_d} \omega^{\frac{k}{2}(k+\varrho_d )} \expect{+_q | k} \expect{k | +_{q'}}, \\ 
&=\frac{1}{d}   \sum_{k\in\mathbb{S}_d}  e^{i \pi (k^2 + k(2(q'-q)+\varrho_d))/d} ,
\label{gauss-ap}
\end{align}
for all $q,q' \in \mathbb{S}_d$ .
This is a generalized quadratic Gauss sum when the GQV is a qudit, and a Gaussian integral when the GQV is a QCV. It can be evaluated using the following two results. For any $a,b \in \mathbb{N}$ such that $a > 0$ and $a+b$ is even then \cite{berndt1981determination}:
\begin{equation} \frac{1}{a} \sum_{k=0}^{a-1} e^{i \pi ( k^2 +b k )/a} = e^{i\frac{\pi}{4}} e^{-i \pi \frac{b^2}{4a}}/\sqrt{a} .
 \end{equation}
As $d \neq 0$ and $d+ 2(q-q') + \varrho_d$ is even ($\varrho_d=0$ and $\varrho_d=1$ for even and odd $d$ respectively), this can be applied to Eq.~\eqref{gauss-ap} for the case of qudits. For QCVs, the following integral relation can be used \cite{watson1928theorems,weisstein2004gaussian}:
\begin{equation}
 \frac{1}{a} \int_{-\infty}^{\infty} \text{d}k \, e^{i \pi ( k^2 +b k )/a} = e^{i \frac{\pi}{4}}  e^{-i \pi \frac{b^2}{4a}}/\sqrt{a} ,
 \end{equation}
which has an exactly equivalent form to the discrete case. Hence, using these two relations it follows that in all cases
\begin{align} \expect {+_q| \times_{q'}} &=  e^{i\frac{\pi}{4}} e^{-i \pi \frac{ (2(q'-q) + \varrho_d)^2}{4d}}/\sqrt{d}, \\
& =\omega^{qq'} \omega^{-\frac{q}{2}(q-\varrho_d)} \omega^{-\frac{q'}{2}(q'+\varrho_d)} \omega^{\frac{d-\varrho_d}{8} } / \sqrt{d} ,
\end{align}
as stated in Eq.~\eqref{overlapconpha} of the main text. 

\vspace{.5cm}
\section{Clifford conjugation relations}\label{App:Clifford}
In this appendix the conjugation relations of the Fourier, phase and $\textsc{cz}$ gates on Pauli operators are derived. It will be shown that 
\begin{align}   
  p_{\xi ,q,q'}   &\xrightarrow{ \mathmakebox[1.2cm]{Z(p)}} p_{\xi+2pq,q,q'}, \label{AEq:conj-Z} \\
  p_{\xi ,q,q'}   &\xrightarrow{ \mathmakebox[1.2cm]{F}} p_{\xi-2qq',-q',q}, \label{AEq:conj-F} \\
 p_{\xi ,q,q'}  &\xrightarrow{  \mathmakebox[1.2cm]{P(p)} } p_{\xi+pq(q+\varrho_d) ,q,q'+pq} , \label{AEq:conj-P} \\ 
p_{\xi ,(q_1 ,q_2 ,q'_1,q'_2)} &\xrightarrow{ \mathmakebox[1.2cm]{\textsc{cz}} }    p_{\xi +2q_1q_2,(q_1,q_2,q'_1+q_2,q'_2+q_1)} \label{AEq:conj-CZ}. \end{align}
as was stated in Eqs.~(\ref{Eq:conj-F} -- \ref{Eq:conj-CZ}) of the main text. The first relation follows directly from the Weyl commutation relation of Eq.~\eqref{Weyl}. That is easily proven, so an explicit proof is not presented.

 We first prove the relation for the Fourier gate. Using Eq.~\eqref{Eq:sum-roots-delta} and the orthogonality relation $\expect{q|q'}=\delta(q-q')$, it follows that
\begin{widetext}
\begin{align}
F Z(p)F^{\dagger} 
&= \frac{1}{d}   \sum_{ p'} \sum_{q,q' }  \sum_{ r,r' }   \omega^{pp'+q q'-rr'} \ket{q}\expect{q' | p' }\expect{p' | r } \bra{r' }, \\
&= \frac{1}{d} \sum_{ p'} \sum_{q,q' }  \sum_{ r,r' } \delta(q'-p') \delta(p'-r) \omega^{pp'+q q'-rr'} \ket{q} \bra{r' },\\
&= \frac{1}{d} \sum_{ p'} \sum_{q }  \sum_{ r' }  \omega^{p'(p+q-r')} \ket{q}\bra{r' },\\
& = \sum_{ q,r '}   \delta(q-r'+p) \ket{q}\bra{r' },\\
& = \sum_{r'  }  \ket{r'-p}\bra{r' },\\
& = X(-p),
\end{align}
\end{widetext}
where all summations are over $\mathbb{S}_d$. Note that, as always, $r'-p$ is to be taken modulo $d$ for a $d$-dimensional qudit. An almost identical derivation implies that  $F X(p)F^{\dagger} = Z(p)$. Then, using the Weyl commutation relation, it follows that
\begin{align*}
 p_{\xi ,q,q'}   
 \xrightarrow{F}  \omega^{\xi/2}  Z(q) X(-q') & =  \omega^{\xi/2} \omega^{-qq'} X(-q') Z(q), \\
&= \omega^{(\xi-2qq')/2} X(-q') Z(q), \\
&= p_{\xi-2qq',-q',q}.
\end{align*}
This confirms the relation claimed in Eq.~\eqref{AEq:conj-F} of this appendix and in the main text. 

Next, consider the relation for phase gate. The conjugation action of the phase gate on $X(q)$ is
\begin{widetext}
 \begin{align}
P(p)X(q)P(p)^{\dagger} & = \sum_{r,s,r }  \omega^{\frac{p}{2} (r(r+\varrho_d) - t(t+\varrho_d)) } | r \rangle \langle r|s+q\rangle \langle s|t\rangle \langle t \rangle, \\
  &= \sum_{r,s,t}\omega^{\frac{p}{2} (r(r+\varrho_d) - t(t+\varrho_d)) }    \delta(s+q-r)\delta(t-s) \ket{r}\bra{t}   , \\
   &= \sum_{r,t} \omega^{\frac{p}{2} (r(r+\varrho_d) - t(t+\varrho_d)) }   \delta(t+q-r)  \ket{r}\bra{t}   , \\
      &= \sum_{t}  \omega^{\frac{p}{2} ((t+q)(t+q+\varrho_d) - t(t+\varrho_d)) }  \ket{t+q}\bra{t}, \label{eqa1} \\
 &= X(q) \sum_{ t}    \omega^{p q(q +\varrho_d )/2 } \omega^{ptq }    \ket{t}\bra{t} ,\label{eqa2} \\
  &=     \omega^{p q(q +\varrho_d )/2 } X(q) Z(pq),
   \end{align}
\end{widetext}
where all summations are over $\mathbb{S}_d$. To get from line \eqref{eqa1} to \eqref{eqa2} the brackets have been expanded. For QCVs, and when $t+q <d$ for qudits, this follows immediately as there is no modulo arithmetic to consider. For qudits, in the parts of the sum where $t+q \geq d$, then $t+q$ represents $t+q-d$ and such a replacement is in general necessary to obtain the correct answer. However, in this case, the calculation with or without this replacement gives the same result (because $\omega$ is $d$ periodic). Because the phase gate commutes with $Z(q)$, it then follows that
  \begin{align} 
    p_{\xi,q,q'}  & \xrightarrow{P(p)}  \omega^{ ( \xi + p q(q +\varrho_d )) /2 } X(q) Z(pq+q') ,\\
& = p_{\xi + p q(q + \varrho_d),q,q'+pq}, 
    \end{align}
 which is the result stated in Eq.~\eqref{AEq:conj-P} and in the main text. 

Finally, consider the $\textsc{cz}$ gate. For a general controlled gate $\C(u)$, defined by
\begin{equation}
\C(u)\ket{q}\ket{q'} = \ket{q}u^{q}\ket{q'},
\end{equation}
the following relation holds:
\begin{equation}
(\mathbb{I} \otimes v^{\dagger} ) \cdot \C(u) \cdot (\mathbb{I} \otimes v ) = \C(v^{\dagger}uv).
\end{equation}
This relation, the Weyl commutation relation, and the equality 
\begin{equation}
\C(\omega^q \mathbb{I} )= Z(q) \otimes \mathbb{I},
\end{equation}
imply that
\begin{equation*} 
X(q_1) \otimes X(q_2)   \xrightarrow{\textsc{cz}}  \omega^{q_1q_2}  X(q_1)Z(q_2) \otimes X(q_1)Z(q_2).
\end{equation*} 
The $\textsc{cz}$ gate commutes with $Z(q)$, and so the above relation implies that
\begin{align} 
p_{\xi ,(q_1 ,q_2 ,q'_1,q'_2)} &\xrightarrow{\textsc{cz} }    p_{\xi +2q_1q_2 ,(q_1,q_2,q'_1+q_2,q'_2+q_1)},
\end{align}
as stated in Eq.~\eqref{AEq:conj-CZ}, and in the main text. 

\section{Generic rotation gates\label{Aunigen}}
Consider the single-qudit gate set
\begin{equation}
\mathcal{G}_{\text{gen}}=\{R(\varphi),F\},
\end{equation}
where $\varphi : \mathbb{Z}(d) \to \mathbb{R}$ is a generic function, where ``generic"  means that $\vartheta(q)$ is uniformly randomly sampled from $[0,2\pi]$ for each $q \in \mathbb{Z}(d)$. By appealing to a standard argument used in Refs.~\cite{lloyd1995almost,deutsch1995universality}, in this appendix it is shown that this gate set can (almost certainly) approximately generate any single-qudit gate. Hence, along with any entangling gate, this gate provides an approximately universal gate set for qudit quantum computation, via the results of \cite{brylinski2002universal}.

If an $R(\vartheta)$ unitary for any $\vartheta:\mathbb{Z}(d) \to \mathbb{R}$ may be approximated to arbitrary accuracy using $F$ and $R(\varphi)$, then these two gates may approximate any single-qudit gate. This follows because Zhou \emph{et al.}~\cite{zhou2003quantum} have shown that any single-qudit unitary can be decomposed into $R(\vartheta)$ and $F$ gates. For a generic function $\varphi : \mathbb{Z}(d) \to \mathbb{R}$ it follows that $\varphi(q)$ and $\varphi(q')$ will be irrational multiples of $\pi$ and each other for every $q,q' \in \mathbb{Z}(d)$ with $q \neq q'$. The intuition behind this is that there are only countably many functions that are not of this sort -- as the rational numbers are countable -- but there are uncountably many functions $\vartheta : \mathbb{Z}(d) \to \mathbb{R}$. For convenience, write these $d$ different phase angles as a vector $\vec{\phi} = (\varphi(0),\dots,\varphi(d-1))$. It is only necessary to be able to generate a rotation gate with any vector of phase angles, $\vec{\theta}$, with the restriction to $\vec{\theta} \in [0,2\pi)^d$, as trivially $e^{i (x +2\pi)}=e^{ix}$. For $N\in \mathbb{N}$, consider
 \begin{align}
 \vec{\phi}_{N} &\equiv N \vec{\phi} \,\,\, \text{mod} \,2\pi ,\\
 &= (N\varphi(0),N\varphi(1),\dots ,N\varphi(d-1) )\,\,\, \text{mod} \,2\pi .
\end{align}
It is known that, for any vector $\vec{\phi}$ with elements that are irrational multiples of $\pi$ and each other, the vectors $\vec{\phi}_1$, $\vec{\phi}_2$, $\vec{\phi}_3$, $\dots$ fill up the interval $[0,2\pi)^d$. Stated another way, the set $\{\vec{\phi}_N \mid N \in \mathbb{N}\}$ is a dense subset of $[0,2\pi)^{d}$. For example, this argument or closely related arguments are made in Refs.~\cite{lloyd1995almost, deutsch1995universality,childs2011characterization}. As such, for a $R(\vartheta)$ gate with any $\vartheta:\mathbb{Z}(d) \to \mathbb{R}$ and given any $\epsilon>0$ there is some $N(\epsilon)\in\mathbb{N}$ such that $R(\varphi)^{N(\epsilon)}$ is an $\epsilon$-approximation to $R(\vartheta)$. A more rigorous proof than that given here could be obtained by adapting the arguments of Ref.~\cite{childs2011characterization}, which are concerned with the universality of two-qubit Hamiltonians and unitaries.

\section{Displacement operators \label{sec:QVs}}
In some contexts the Pauli operators are replaced by the entirely equivalent \emph{displacement operators}. This is particularly common with QCVs in the setting of quantum optics \cite{gerry2005introductory,radmore1997methods}, but sometimes the theory of qudits is also presented in terms of these operators \cite{saraceno1990classical,marchiolli2007discrete,klimov2009discrete}. The displacement operators for a GQV may be defined by their relation to the Pauli operators \footnote{In this definition, $2^{-1}$ is the multiplicative inverse of 2 in $\mathbb{S}_d$. For QCVs this is obviously $1/2$. For odd dimension qudits this always exists ($d$ and 2 are co-prime) and is $(d+1)/2$. For even dimensions this phase factor could be omitted, or $2^{-1}$ could be replaced with $1/2$.}
\begin{equation}
\mathcal{D}(q,q'):=\omega^{-2^{-1} qq'}  Z(q') X(q).
\end{equation}
The relationship between Pauli operators, the displacement operators, and coherent states is briefly discussed in this appendix.

The standard definition of the complex-number parameterized displacement operator is
\begin{equation}
\mathcal{D}(\alpha) \equiv \mathcal{D}\left(\sqrt{2} \Re (\alpha),\sqrt{2} \Im (\alpha)\right),
\label{C-real-disp}
\end{equation}
where $ \Re(\alpha)$ and $\Im(\alpha)$ are the real and imaginary components to $\alpha \in \mathbb{C}$, respectively.
For a QCV (and particular in the setting of optics) it is conventional to express the displacement operator in what is termed the `entangled' form
\begin{equation}
\mathcal{D}(q,q')= \exp (i(q' \hat{x} - q \hat{p})).
\label{displace}
\end{equation}
This may be derived from the relations $Z(q)=\exp(iq \hat{q})$ and $X(q)=\exp(-iq\hat{p})$ along with the canonical commutation relation and the Weyl formula 
\begin{equation}
e^A e^B = e^{\frac{1}{2}[A,B]} e^{A+B} ,
\label{Weyl-formula}
\end{equation}
that holds when $[A,[A,B]] = [B,[A,B]]=0$, and which is a special case of the Baker-Campbell-Hausdorff formula \cite{gazeau2009coherent}. The QCV displacement operator is also often written in terms of the \emph{creation} and \emph{annihilation} operators
\begin{equation}
\hat{a}^{\dagger} := \frac{1}{\sqrt{2}} (\hat{x} - i \hat{p}), \hspace{1cm} \hat{a} := \frac{1}{\sqrt{2}} (\hat{x} + i \hat{p}),
\label{axprel} 
\end{equation}
which obey the commutation relation $[\hat{a},\hat{a}^{\dagger}]=1$. It is easily shown that
\begin{equation}
\mathcal{D}(\alpha) = \exp(\alpha a^{\dagger} - \alpha^{*} a),
\label{complexdi2}
\end{equation}
with $\alpha \in \mathbb{C}$. This form is the most common in quantum optics. From this equation, which is often used to \emph{define} the displacement operator, it is certainly not obvious (at least to me) that this operator is analogous to the qubit Pauli operators.

Displacement operators may be used to define the $\mathbb{C}$-number parameterized coherent states by
\begin{equation}
 \ket{\alpha } := \mathcal{D}(\alpha) \ket{\psi_0},
\label{cohst}
\end{equation}
where $\ket{\psi_0}$ is some reference state. The well-known Glauber (or standard) coherent states \cite{glauber1963coherent} are obtained for a QCV with the reference state as the vacuum, $\ket{\text{vac}}$, which is the lowest energy eigenstate of the quantum Harmonic oscillator. For qudits, coherent states are less often considered but one choice of reference state to define them is an eigenstate of the Fourier transform $F$ \cite{klimov2009discrete}.
\end{document}